*Pomazanov M.V.*

# *The connection of the stability of the binary choice model with its discriminatory power*

**Abstract**

The key indicators of model stability are the population stability index (PSI), which uses the difference in population distribution, and the Kolmogorov-Smirnov statistic (KS) between two distributions. When deriving a binary choice model, the question arises about the real Gini index for any new model. The paper shows that when the Gini changes, the real Gini index should be less than the obtained Gini index. This type is included in the equation using a formula, and the PSI formula in KS is also included based on the scoring indicator. The error in calculating the Gini index of the equation is unavoidable, so it is necessary to always rely on the calculation formula. This type of research is suitable for a wide range of tasks where it is necessary to consider the error in scoring the indicator at any length.

*Помазанов М.В.*

## *Связь стабильности модели бинарного выбора с ее дискриминирующей мощностью*

**Abstract**

Ключевые показатели стабильности модели считаются так - показатель стабильности популяции (PSI), использующий разность в распределении популяций, а также - содержательная мера куда включается статистика Колмогоров-Смирнов (KS) между двумя распределениями. При выводе модели бинарного выбора возникает вопрос о реальном индексе Джини для любой новой модели. В работе показано, что при изменении Джини реальный индекс Джини должен быть меньше, чем полученный. Данный тип вносится формулой в уравнение, вносится также и формула учета PSI в KS на основе scoring indicator. Ошибка пересчета погрешности индекса Джини уравнения является не устранимой, поэтому необходимо всегда опираться на формулу пересчета. Представленный тип исследования пойдет для большого количества задач, где необходимо учитывать погрешность скоринга индикатора на любой длине.

.

## 1. Introduction

Ключевым показателем стабильности скоринговой модели считается показатель стабильности популяции (PSI). В частности, при моделировании кредитного риска PSI является наиболее широко используемым показателем для мониторинга эволюции популяции, лежащей в основе модели, путем оценки степени расхождения или наоборот, сходства между двумя дискретными распределениями вероятностей – базовой и новой,  (Thomas et al., 2002, pp. 155 ff.) and (Siddiqi, 2017,pp. 368 ff.)).

Другие области применения индекса стабильности включают  страхование, здравоохранение, инженерия и маркетинг  (Huang et al. , 2022), (Li et al. , 2022), (Sahu et al., 2023), (Dong et al., 2022), (Wu & Olson, 2010), (McAdams et al. , 2022), (Chou et al. , 2022), (Karakoulas, 2004) and (Brockett et al., 1995).

Дискретная формула для показателя стабильности представляется в виде (see Fig. 1 right) $PSI = \sum_{i \in buckets}(p_i - p_i^N) \cdot ln\left(\frac{p_i}{p_i^N}\right)$

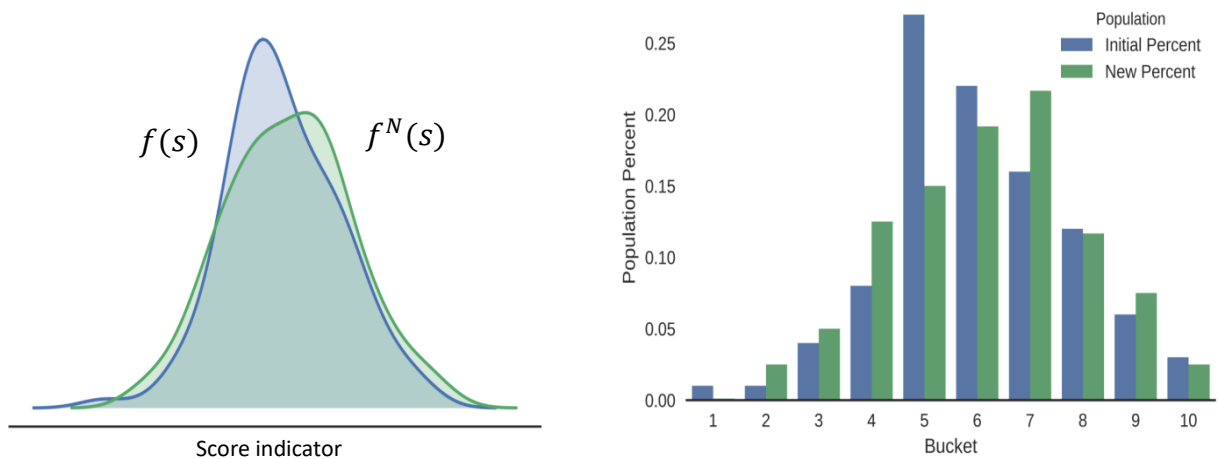

*Figure 1. Непрерывный и дискретный аналог распределений score indicator для начальной и новой популяции*

Непрерывный аналог PSI при $f(s) > 0, f^N(s) > 0$ определяется по формуле в виде (see Fig. 1 left)

$$PSI = \int_{-\infty}^{+\infty}(f(s) - f^N(s)) \cdot \ln\left(\frac{f(s)}{f^N(s)}\right)ds \qquad (1)$$

Показатель PSI используется в регуляторных целях валидации качества моделей оценки кредитного риска, когда модель используется для оценки требований к капиталу (European Central Bank, 2024, 2019), (Board of Governors of the Federal Reserve System, 2011), (South African Reserve Bank, 2022). Российским регулятором четко установлена недопустимая зона показателя $PSI > 0.25$ (Bank of Russia, 2024).

Другая содержательная мера включается в статистику Колмогоров-Смирнов

(D'Agostino & Stephens, 1986) $KS = \max_{i \in buckets} |F_i - F_i^N|$, где $F_i = \sum_{k \leq i} p_i$, $F_i^N = \sum_{k \leq i} p_i^N$. Которая так же имеет непрерывный аналог (see Fig.2):

$$KS = \max_s \left| \int_{-\infty}^{s} (f(x) - f^N(x)) dx \right| \qquad (2)$$

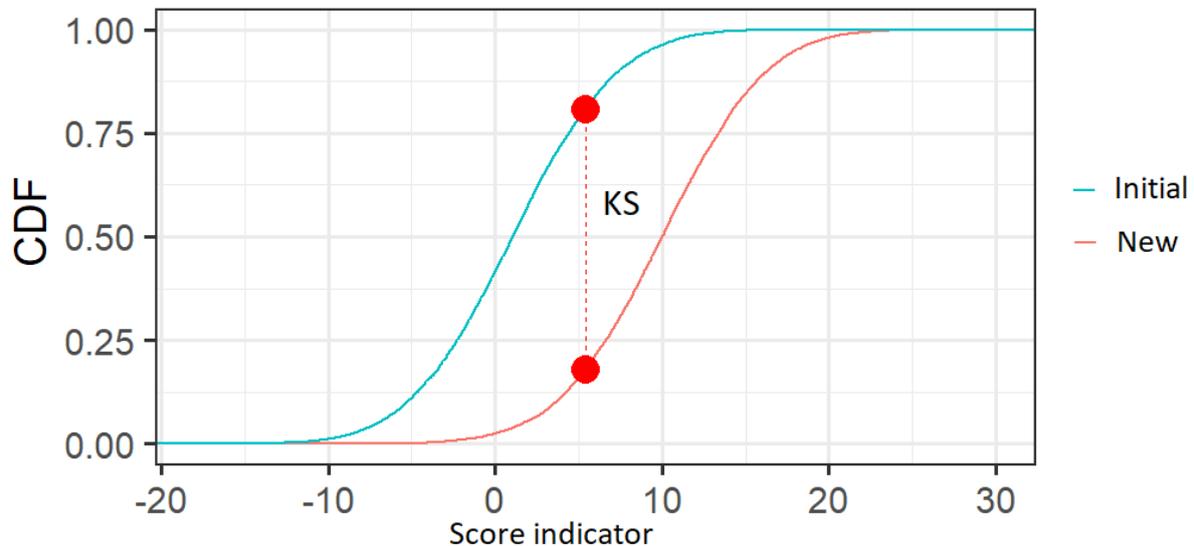

*Figure 2. Cumulative distribution functions and KS-metric*

Метрика KS имеет очевидный математический смысл в оценке степени расхождения двух распределений. А именно, в каком значении кумулятивной доли окажется фиксированный score indicator в новом распределении score относительно первоначального, при условии, что score indicator фиксируется в точке максимального расхождения.

Дискриминирующая мощность модели определяется с помощью ROC-кривой. Предположим, что все «плохие» и «хорошие»[1] наблюдения упорядочены в соответствии с присвоенным скоринговым баллом от худшего к лучшему. Тогда точки ROC кривой определяются как соответствие доли «хороших» наблюдений доле «плохих» наблюдений, в которой присвоенные скоринговые баллы хуже, чем лучший скоринговый балл из первой группы. ROC-кривая (Provost & Fawcett, 2001) — это непараметрический инструмент оценки эффективности, который представляет собой компромисс между истинно положительным (TP) и ложноположительным (FP) процентами классификаций примеров на основе непрерывного вывода по всем возможным значениям порога принятия решения (see Fig.3)

---

[1] «Плохое» наблюдение – это наблюдение, которое испытало отрицательное событие согласно данному в постановке бинарного выбора определению. Например, дефолт заемщика определяется как факт выявление просрочки более 90 дней, отказ выполнения обязательств, судебное решение в пользу кредитора и т. д. «Хорошее» наблюдение – наблюдение, не являющееся «плохим».

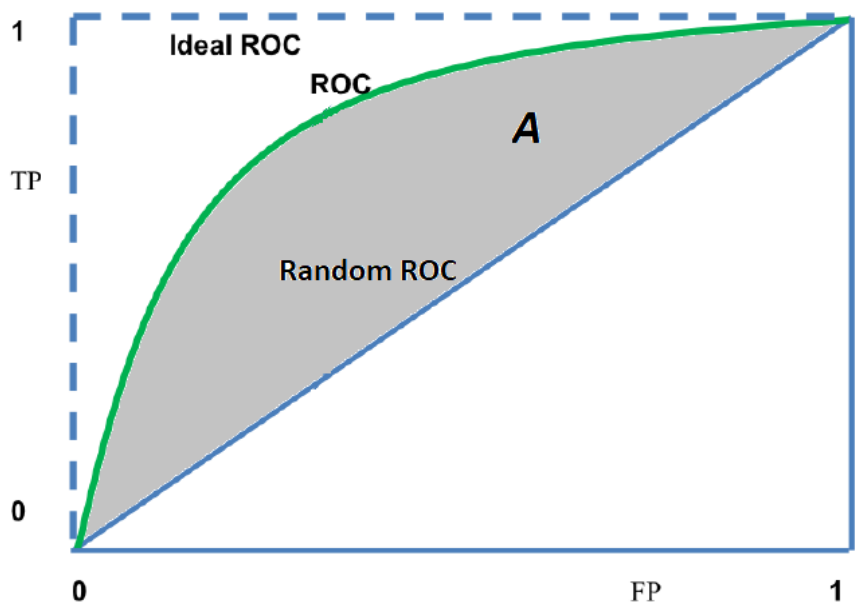

*Figure 3. ROC-curve*

Широко используемой метрикой качества скоринговой модели является дискриминирующая мощность, связанная с площадью под ROC-curve,

$\text{AUROC} = \frac{1}{2} + \text{A}$ или индекс Джини $\text{G} = 2 \cdot \text{A}$. Индекс G находится в диапазоне от 0 до 1, 0 – означает, что скоринговая модель не лучше, чем случайная, 1 – означает, что модель идеальна и на 100% предсказывает верный результат. Более подробно в статье (Engelmann et al., 2003) анализируется кумулятивный профиль точности (CAP) и операционную характеристику получателя (ROC), предлагается теоретико-тестовая интерпретация вогнутости CAP и ROC-кривой и показывается, как это наблюдение может быть использовано для более эффективного использования информационного содержания, в том числе: как выявить факторы с высокой дискриминационной способностью, как рассчитать доверительные интервалы для площади под ROC-кривой и как проверить, отличаются ли две скоринговые модели, валидированные на одном наборе данных.

Стандартное отклонение статистической погрешности индекса Джини при условии $G \geq 0$ вычисляется по формуле (Hanley&McNeil, 1982)

$$\sigma_G = \sqrt{\frac{1 - G^2 + (N_b - 1) \cdot \left(4 \cdot \frac{G+1}{3-G} - (G+1)^2\right) + (N_g - 1) \cdot \left(4 \cdot \frac{(G+1)^2}{3+G} - (G+1)^2\right)}{N_b \cdot N_g}} \quad (3)$$

которая является асимптотической по $N_g, N_b$ где $N_g$ – количество «хороших» наблюдений, $N_b$ – количество «плохих» наблюдений. В кредитном скоринге наблюдения проводятся не чаще одного раза в год по одному заемщику, чтобы обеспечить их приблизительную независимость. В идеале, наблюдения надо проводить один раз по одному объекту, формула (3) была получена для проблем медицины, где наблюдения больных независимы.

В представленной работе будет исследоваться вопрос, как нестабильность популяции (т.е. нестабильность распределения score indicator) будет отражаться на эффективной дискриминирующей способности модели, которая будет наблюдаться на практике при установлении порога принятия решения. Интуитивно понятно, что с ростом нестабильности популяции (увеличение PSI) погрешность эффективной дискриминирующей способности (эффективный индекс Джини) должна увеличиваться относительно измеренной на прошлых данных. Исследуется насколько эта погрешность сопоставима со статистической погрешностью (3), учитывающей ограниченность количества наблюдений в сегменте модели.

До настоящего времени многократно исследовался вопрос о связи между нестабильностью рейтингов и точностью бинарного прогнозирования. В работе (Cantor & Mann, 2007) авторы от лица кредитного рейтингового агентства (CRA) Moody's показали, что более редкий пересмотр рейтингов, снижающий волатильность, приведет к ухудшению дискриминирующей мощности. В работе (Carvalho et al., 2014) обнаружено, что интенсивность, с которой CRA меняют рейтинги, меняется с течением времени, но периоды более сильных изменений рейтингов не связаны с более высокой точностью рейтингов. Аналогично, авторы (Kiff & Kisser, 2024), используя регрессионный анализ, не смогли найти эмпирических доказательств того, что волатильные рейтинги в среднем более точны. Этот вывод устойчив во времени и сохраняется в различных классах активов.

Проблема, которая исследуется в настоящей работе, ставит тот же вопрос о точности прогнозирования, но для скоринговой модели, которая имеет показатель точности, индекс Джини, измеренный на прошлых наблюдениях. Однако, исследуемая точность прогнозирования рассматривается в условиях фактической ограниченной стабильности популяции, измеренной показателем PSI, RS.

## 2. Research Methodology

Допустим мы имеем две скоринговые модели с индексами Джини $G_1$ and $G_2$, пусть $G_1 > G_2$. Считаем, что Модель 2 абсолютно устойчива и распределение scoring indicator от года к году не меняется, метрика Колмогоров-Смирнов $KS = 0$, для Модели 1 наблюдается неустойчивость, при которой $KS = \Delta > 0$. Далее допускаем, что изменение от года к году распределений scoring indicator наблюдений «хороших» и «плохих» имеют идентичную природу, т. е. сдвиги одинаковы. В случае, если $N_g \gg N_b$ этим допущением даже можно пренебречь.

На Рис. 4 изображены две ROC-curve моделей 1 и 2, обладающие дискриминирующей мощностью $G_1$ и $G_2$ соответственно. Поскольку стабильность модели 2 является абсолютной, то уровень принятия решения по скорингу $s_C$ будет обеспечивать расчетный уровень отклонения «плохих» для этой модели. Однако, отклонение распределения для модели 2, после установления уровня CutOff на основе прошлых наблюдений, может произойти в неблагоприятную сторону такую, что доля отклоненных «плохих» для более мощной модели 1 будет идентична отклонению той же доли для менее мощной модели 2.

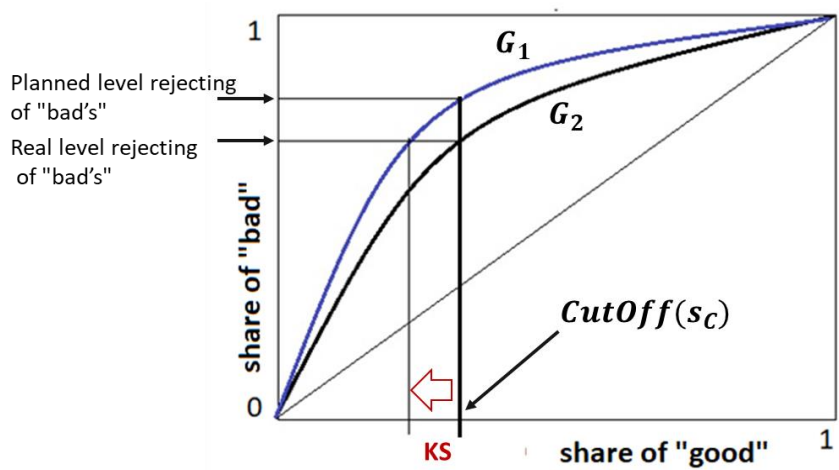

*Figure 4. Возможный сдвиг доли отвержения «плохих» после отклонения распределения score indicator при постоянном decision threshold $s_C$.*

Для практики это будет означать снижения эффективности дискриминирующей мощности $G_1$ обусловленной наличием нестабильности. Проведем консервативные оценки того, насколько может быть снижен эффективный индекс Джини $G_2 = G_{Low}$ относительно измеренного на прошлых испытаниях индекса $G_1 = G$. Консервативность оценки означает, что решение по CutOff принятое на уровне $s_C$ близко к точке, где достигает максимума функция метрики KS (2). Т. е. сдвиг распределения произошел на уровне Δ, а значит

$$ROC_{G_{Low}}(s_C) = ROC_G(s_C - \Delta) \qquad (4)$$

Для оценки соотношения между $G_{Low}$ и $G$ достаточно воспользоваться первым членом разложения соотношения в ряд Тейлора по значению Δ. А также предположим, что вид ROC-curve определяется гармонической однопараметрической симметричной функцией типа (Pomazanov, 2021)

$$ROC_\beta(x) = \frac{(1+\beta)x}{x+\beta} \qquad (5)$$

индекс Джини $G(\beta) = 2 \cdot (1+\beta)\left(1 - \beta \cdot ln\left(1 + \frac{1}{\beta}\right)\right) - 1, \beta \in (0, \infty), \lim_{\beta \to 0} G(\beta) = 1$.

Функция $G(\beta)$ монотонно убывающая, поэтому, если $G(\beta) = G$, то индекс Джини $G_E$ параметризуется как $G_{Low} = G(\beta + \delta) < G, \delta > 0$. Тогда, для переменного уровня $x$ принятия решения получим уравнение (4) в виде

$$\frac{(1+\beta)(x-\Delta)}{x-\Delta+\beta} = \frac{(1+\beta+\delta)x}{x+\beta+\delta} \tag{6}$$

Выражая $\delta(x,\beta)$ из (6) и исходя из принципа консервативности $\delta \to max$, который достигается при $x^* = \frac{1+\Delta}{2}$, получается $\delta_\beta(\Delta) = \frac{4\beta\Delta(1+\beta)}{(1-\Delta)^2 - 4\beta\Delta}$. Уровень $x^*$ может быть близок к уровню принятия решения $CutOff(s_C)$. Опуская члены более высокого порядка малости, чем $\Delta^1$, после разложения $G(\beta + \delta_\beta(\Delta))$ в ряд Тейлора, получаем

$$\begin{gathered} G_{Low} = G - \Delta \cdot \Omega(\beta), \\ \Omega(\beta) = 8\beta \cdot (1+\beta) \cdot \left((1+2\beta) \cdot ln\left(1+\frac{1}{\beta}\right) - 2\right) \end{gathered} \tag{7}$$

Из-за трансцендентности зависимости $G(\beta)$ (5) аналитическую формулу $\Omega(\beta) = \omega(G)$ (7) построить не удается. Однако, оказывается возможно построить очень близкую аппроксимацию $\omega(G)$ в виде $\omega(G) \cong \omega_A(G) = \Omega_0 \cdot (1 - G^\gamma)$,
где $\Omega_0 = 1.323, \gamma = 2.204$ (смотри рис.5).

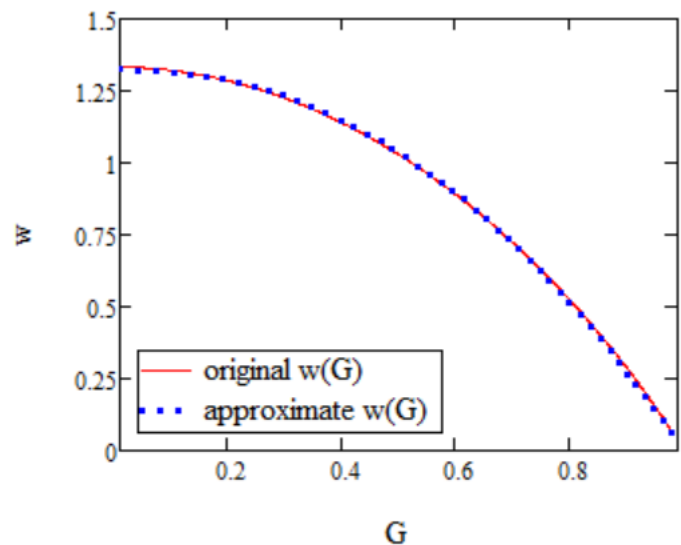

*Figure 5. Точная зависимость $\omega(G)$ и ее аппроксимация*

Нестабильность популяции может проявиться и в сторону увеличения эффективного индекса Джини, тогда возможно не оптимальное отклонение избыточного числа «хороших» наблюдений (ошибка второго рода), однако на практике максимальные издержки происходят при ошибке первого рода (менее чем нужно отклонить «плохих»), поэтому разумно констатировать оценку погрешности индекса Джини, связанную с нестабильностью популяции, как

$$\Delta G = \Delta \cdot 1.3 \cdot (1 - G^{2.2}) \tag{7}$$

где $\Delta$ – средняя метрика Колмогоров-Смирнов сдвига распределений популяции по результатам наблюдений год к году, $G$ – метрика сдвига.

## 3. The relationship of stability metrics

Рассмотрим, как связаны метрики KS и PSI при разумных допущениях. Пусть

$$f_\lambda(x) = f(x) \cdot (1 + \lambda \cdot \delta(x)), \lambda \ll 1 \qquad (8)$$

тогда из $\int_{-\infty}^{+\infty} f(x)dx = 1$, $\int_{-\infty}^{+\infty} f_\lambda(x)dx = 1$, будет следовать $\int_{-\infty}^{+\infty} f(x)\delta(x)dx = 0$. Откуда, при единственности функции $\delta(x) = 0$,

$\delta(x) > 0$, при $x_0 < 0$, получаем из (2)

$$KS = \lambda \cdot \max_{s \in (-\infty, +\infty)} \int_{-\infty}^{s} f(x)\delta(x)dx = \lambda \cdot \int_{-\infty}^{x_0} f\delta dx$$

С другой стороны, из (1)

$PSI = \lambda \cdot \int_{-\infty}^{+\infty} f(x)\delta(x) \cdot \ln(1 + \lambda \cdot \delta(x))dx = \lambda^2 \int_{-\infty}^{+\infty} f\delta^2 \, dx + O(\lambda^3)$. Значит имеем разумное допущение, что

$$KS = \sqrt{PSI} \cdot \frac{\int_{-\infty}^{x_0} f\delta dx}{\sqrt{\int_{-\infty}^{+\infty} f\delta^2 \, dx}} \cdot (1 + O(\lambda)) \qquad (9)$$

Давайте обоснуем предположение (9) на примере совокупности отсчётов за 2023 год (Moody's, 2024). В последней таблице Moody's даны матрицы переходов в нулевые состояния, отложим только их и дадим формулы матриц переходов из года в год по всем состояниям. Матрицы даны состоянием в каждой точке на каждом году деятельности 1971-2023. Матрица деятельности буде вычисляться по статистике Таб. 1

|       | 1970 | 1971 | 1973 | ... | 2021 | 2022 | 2023 |
|-------|------|------|------|-----|------|------|------|
| Aaa   | 39   | 40   | 41   | ... | 50   | 50   | 50   |
| Aa    | 77   | 74   | 81   | ... | 330  | 317  | 319  |
| A     | 254  | 282  | 309  | ... | 1289 | 1321 | 1329 |
| Baa   | 372  | 398  | 441  | ... | 1863 | 1927 | 1860 |
| Ba    | 238  | 228  | 205  | ... | 780  | 772  | 699  |
| B     | 36   | 27   | 27   | ... | 891  | 955  | 866  |
| Caa-C | 16   | 8    | 6    | ... | 1460 | 1589 | 1525 |

*Table 1. Распределение матриц*

После построения матриц по статистикам (1) и (2) будем иметь соотношения на следующем Рис. 6

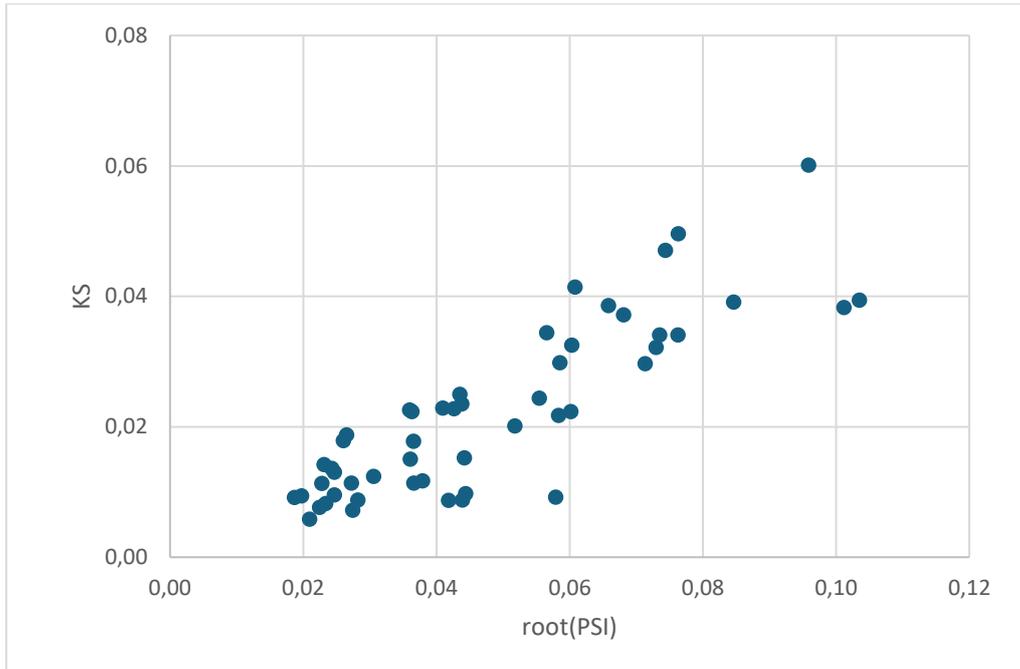

*Figure 6. Зависимость между PSI и КС в фигуре матрицы отчетов за 1971-2023*

Видно, что соотношение между КС и PSI более-менее соблюдаются, хотя нечетко и не равно постоянному числу в матрице (9), равному примерно 2/5.

Рассмотрим модель, которая определяется согласно $Q = \frac{\int_{-\infty}^{x_0} f\delta dx}{\sqrt{\int_{-\infty}^{+\infty} f\delta^2 dx}} \cdot (1 + O(\lambda))$.

Предположим, модель имеет $Q = 0.5$, тогда нижняя граница индекса Джини опишется прямыми (7), см. Рис. 7.

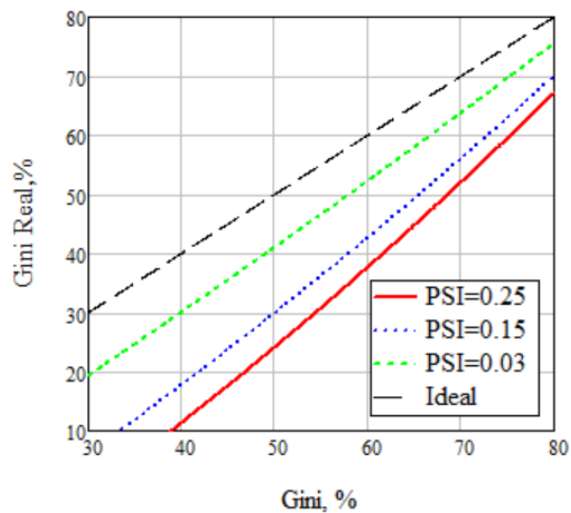

*Figure 7. Статистика для Q=0.5*

Из Fig.7 видно, что для преодоления Gini Real > 50% для Gini=70% при $\gamma = 0.25$, т. е. потребуется нижняя кривая. Очевидно, видно и другое, что для легкого снижения Джини – требуется изначально Джини выше (т. е. Ideal).

4. Discussion

Для передачи устойчивого Джини для оценки модели требуется, чтобы погрешность модели была не ниже (7), однако это требует, чтобы модель была устойчива. Поэтому, для введения параметра неустойчивости модели $\Delta G$ требуется задание начального $G, \Delta$. Это ничего не имеет общего с заданием параметров $N_g, N_b$ для формулы (3), которая дает стандартное отклонение для $G$ при равных $N_g, N_b$. При этом, $\Delta G \to 0$ при $N_g \to \infty, N_b \to \infty$, что не так для формулы (7).

Формула, представленная уравнением похожим на (7), дает $\Delta G$ близкое к тому, что будет реализовано. При выводе (7) мы брали максимум два раза, однако на практике может оказаться, что решение близко к максимуму, поскольку самые крутые решения часто сидят именно на максимумах. Устойчивость (7) смотреть скучно, поскольку не видно изначально близко к каким максимумам будет производиться решение. Поэтому мы будем оптимизировать близко к текущим максимумам, т. е. формулой (7).

Предположим, что существует решение близкое к (7), при $Q = 0.4$ и $PSI = 0.1$, то $\Delta G = 0.16 \cdot (1 - G^{2.2})$, т.е. при $G = 60\%$, получается $\Delta G = 10\%$, однако, при $PSI = 0.01, \Delta G = 4.8\%$. Тогда, решение по снижению разброса индекса Джини должно быть на уровне $PSI = 0.01$. Это означает, что снижение разброса должно исходить с уровня $PSI = 0.1$ до уровня $PSI = 0.01$. Это чуть более, чем в два раза.

С другой стороны, пусть в процессе изменения системы у нас складывается ситуация, что погрешность с учетом системных изменений не ниже той, которая была бы при не изменении координат. Тогда, очевидно, никаких изменений делать в системе не нужно. Изменения производятся только при складывании так, чтобы погрешность с учетом системных изменений была бы выше той, которая обеспечивалась неизменностью.

5. Conclusion

При определении изменений в систему, после соответствующих изменении координат, возникает вопрос, когда их делать, а когда нет. Понятно, что при изменениях, накладывающих существенные коррекции в систему, изменять надо, но при условии, что изменения больше, чем требуется. В представленной статье мы привели требования такие, что при выпуске обновленной системы изменения принимать нужно. Продемонстрировали, что формула (7) работает при достаточно низком уровне изменений в распределение полярности. Одновременно показали, что изменения (7) плохо работают, когда будущее существенно определяется текущей величиной.

Когда выводилась (7) бралось двойное поляризационное соотношение, что показывает, что (7) в общем то груба, но реальные правки, оказывается, ненамного меньше (7). Формула (7) не является никакой языковой загогулиной, типа (3), поскольку не изменяет Джини при увеличении количества «хороших» и «плохих», она является свободой изменения распределений. Поэтому, формула типа (7) должна применяться при разных изменениях определения системы, но существенно опускать индекс Джини необходимо при обоснованном пересмотре.

Мы неоднократно проводили пересмотр системы, но всегда нарывались на соотношения типа (9), хотя не столь однозначны, как даны. Однако всегда видели существенное отличие от первоначальной формулы. Т. е. формулу надо применять даже, если дано только PSI. Это оправдано по большей части тем, что PSI тоже дает высокое отличие от первоначальной формулы. В любом случае, необходимо смотреть на KS двух отличающихся распределений.

Представленный тип исследования пойдет для большого количества задач, где необходимо учитывать погрешность скоринга индикатора на любой длине.